\documentclass[preprint,prd,nofootinbib,onecolumn,showpacs]{revtex4}
\usepackage{dcolumn}
\usepackage{lipsum}
\usepackage{graphicx,epsfig}
\usepackage{psfrag}
\usepackage{bm}
\usepackage{epstopdf}
\usepackage{latexsym}
\usepackage{multirow}
\usepackage{amsmath,amssymb}
\usepackage{enumerate}
\usepackage{hyperref}
\usepackage[FIGTOPCAP]{subfigure}

\def\be {\begin{equation}}
\def\ee {\end{equation}}
\def\ba {\begin{eqnarray}}
\def\ea {\end{eqnarray}}
\def\nn {\nonumber}
\def\bc {\begin{center}}
\def\ec {\end{center}}
\newcommand{\bdm}{\begin{displaymath}}
\newcommand{\edm}{\end{displaymath}}

\usepackage{multirow}
\usepackage{amsmath,amssymb}
\usepackage{enumerate}
\usepackage{hyperref}

\def\be {\begin{equation}}
\def\ee {\end{equation}}
\def\ba {\begin{eqnarray}}
\def\ea {\end{eqnarray}}
\def\bc {\begin{center}}
\def\ec {\end{center}}
\def\bi {\begin{itemize}}
\def\ei {\end{itemize}}

\def\a  {\alpha}

\def\da {\dagger}

\def\f {\frac}

\def\g  {\gamma}

\def\k  {\kappa}

\def\la {\label}
\def\le {\left}

\def\na {\nabla}
\def\nn {\nonumber}

\def\O  {\Omega}

\def\pa {\partial}

\def\ra {\rightarrow}
\def\ri {\right}

\def\s {\sigma}

\def\sq {\sqrt}
\def\t  {\tau}

\def\> {\rangle}
\def\< {\langle}

\def\hs {\hspace}
\begin{document}

\title{Renormalized stress tensor of a quantized massless scalar field in warped cosmological braneworld background} 

\author{Suman Ghosh\footnote{Electronic address : {\em suman.ghosh@igntu.ac.in}}${}^{}$}
\affiliation{Department of Physics \\
Indira Gandhi National Tribal University, Amarkantak, MP - 484887, India}

%\author{Sayan Kar\footnote{Electronic address : {\em sayan@cts.iitkgp.ernet.in}}${}^{}$}
%\affiliation{Department of Physics and Centre for Theoretical Studies
%\\ Indian Institute of Technology Kharagpur, WB - 721 302, India}

\begin{abstract}
Energy momentum tensor of a conformally coupled quantum scalar field in five dimensional warped cosmological spacetimes is studied. We look at situations where the four dimensional part represents a cosmological thick brane and the scale of the extra dimension is time dependent. 
Renormalization of the components of the energy momentum tensor is achieved using adiabatic regularization method. The resulting energy and pressure densities explicitly show the effects of warping and the dynamic extra dimension on the created matter. We discussed how the created matter may accumulate to form thick branes along the extra dimension. % 

%The warp factor is chosen to be that of the Randall--Sundrum model. With particular choices for the functional form of the scale factor (and also the function characterising the time evolution of the extra dimension) we obtain the ${\vert \beta_k\vert}^2$, the particle number and energy densities after solving (wherever possible, analytically but, otherwise, numerically) the conformal scalar field equations. The behaviour of these quantities for the massless and massive Kaluza--Klein modes are examined. Our results show the effect of a warped extra dimension on particle creation and illustrate how the nature of particle production on the brane depends on the nature of warping, type of cosmological evolution as well as the temporal evolution of the extra dimension.    
\end{abstract}

%----------------------------------------------------------------------%
\pacs{04.62.+v, 04.50.-h, 11.10.Kk}

\maketitle

%---------------------------------------------------------------------%

\section{Introduction}

Since Kaluza-Klein \cite{Kaluza,Klein:1926tv}, a large variety of cosmological models with extra dimensions have been proposed over the years. The braneworld models \cite{Randall:1999ee, Randall:1999vf,Gogberashvili:1998vx, ArkaniHamed:1999hk} where our world is considered as a four dimensional sub-manifold (3-brane) embedded in five dimensions are popular today largely because of their potential in proposing achievable experimental signatures of extra dimensions.  
The warped type assumes a curved higher dimensional spacetime and the line element on the 3--brane is {\em scaled} by a {\em warp factor}, thus rendering the higher dimensional metric non-factorisable. 
Brane-world models seem to provide a viable resolution of the  long-standing hierarchy problem in high energy physics. How a quantized field behave in such background is therefore an important question.
Analysis of quantum fields in a higher dimensional spacetime with a Kaluza--Klein-like extra dimension has been reported in  \cite{RandjbarDaemi:2000cr,Saharian:2005xf,Garriga:1989jx,Nojiri:2003vk,Mak:1999tf,Huang:1989zy}. 
Gravitational particle production in braneworld cosmology and its implications have been studied in \cite{Bambi:2007nz}. However, investigations on quantum fields have mostly done in time-independent braneworld backgrounds which do not take into account the expansion of the 3-brane as such. Interestingly large class of warped cosmological i.e. time dependent bulk solutions in presence of various bulk fields were found e.g. in \cite{Ghosh:2008vc}. In  \cite{Ghosh:2008zs}, the author have studied how such various kind of warped and dynamic extra dimensions effect the particle creation rate of a massless bulk scalar field. 

Quantum field theory in curved spacetime \cite{PT,MW,Wald-QFTCS,Buchbinder,Fulling,BD} is an useful tool to study particle creation by fields due to evolving geometry of spacetime. It has also been instrumental to analyse the inhomogeneities in the cosmic microwave background and the large-scale structure of the Universe \cite{Liddle:2000cg,Hawking:1974sw}. The concept of the so-called adiabatic vacuum gives us a notion of particles in curved space that comes closest to the definition of field quanta in flat space. One important feature of the energy momentum tensor (EMT) of the created particles in curved spacetime are the quadratic and logarithmic ultra-violet (UV) divergences in addition to the expected quartic divergence.
Various renormalization methods were developed to tame these infinities. The one we are going to use in this article is adiabatic regularization \cite{parker:2012,Parker:1974qw,Zeldovich:1971mw}. In this method, finite expressions are obtained from the formal one containing UV divergences by subtracting mode by mode (under the integral sign) each term in the adiabatic expansion of the integrand which contains at least one UV divergent part for arbitrary values of the parameters in the theory.

A study of quantum scalar field and renormalization in the context of warped spacetimes where there is a dynamic extra dimension is our objective here. Recent discovery of Higgs particle also motivates us with the following investigation.
The plan of the article is as follows. In Section II, we review the conformally coupled scalar field equations and the adiabatic regularization method to find the renormalized EMT (REMT). Here we have essentially used the algorithm introduced in \cite{Zeldovich:1971mw}. We write the equations to determine particle number density in terms of, which may be called, Zeldovich-Starobinsky (ZS) variables that were introduced in \cite{Zeldovich:1971mw}. The equations for these ZS variables can be solved as adiabatic series where each term in the series are essentially functions of metric functions and their time derivatives.
In Section III, we derive the REMT using the adiabatic subtraction method. We emphasize on how components of REMT depend on metric functions to figure out the distinguishing roles of the warping factor and the time-dependent scale factors as such. Apart from the adiabatic regularization we discuss how this created matter density may be localised along the extra dimension, thus providing a notion of a {\em physical} thick brane.
Finally, in Section IV, we conclude with comments and future plans.

%Higgs as a scalar field.

\section{Quantum field coupled to a spacetime with an extra dimension}  

Let us consider the background line element (using conformal time) to be generically of the form \cite{Ghosh:2008vc}
\begin{equation} 
ds^2 = e^{2f(\sigma)} a^2(\eta)[- d\eta^2 + dx^2 + dy^2 + dz^2] + b^2(\eta) d\sigma^2 ,\label{eq:metric}
\end{equation}
where $ e^{2f(\sigma)}$ is the warp factor, $a(\eta)$ and $b(\eta)$ are the scale factors associated with the ordinary space (${\vec x}$) and the extra dimension ($\sigma$) respectively. $\eta$ denotes conformal time.
 
The Lagrangian density of a scalar field $\psi(\eta,\vec{x},\sigma)$ coupled to the background geometry is given by \cite{BD}
\be
{\cal L} = - \f{1}{2}\le[g^{AB}\pa_A\psi\pa_B\psi + (m^2 + \xi R)\psi^2\ri], ~~~~ A = 0,1,2,3,4. \la{eq:lag}
\ee
In case of conformal coupling in $d$ spacetime dimensions, the coupling constant $\xi = \f{d-2}{4(d-1)}$. The corresponding momenta is given by
\be
\pi = \f{\pa \sqrt{-g}{\cal L}}{\pa \dot \psi} = a^2 e^{2f} \dot \psi.
\ee
Eq. (\ref{eq:lag}) leads to the following Klein-Gordon equation,
\begin{eqnarray}  
& & -\ \frac{e^{-2f}}{a^2} \frac{\partial^2 \psi}{\partial \eta^2} + \frac{e^{-2f}}{a^2} \left(\frac{\partial^2 \psi}{\partial x^2} + \frac{\partial^2 \psi}{\partial y^2} + \frac{\partial^2 \psi}{\partial z^2}\right) + \frac{1}{b^2} \frac{\partial^2 \psi}{\partial \sigma^2}\nonumber \\ 
& & -\ \frac{e^{-2f}}{a^2} \left (\frac{2\dot a}{a} + \frac{\dot b}{b} \right )\frac{\partial \psi}{\partial \eta} + \frac{4f'}{b^2} \frac{\partial \psi}{\partial \sigma} - (m^2 + \xi R)\psi = 0 ,\label{eq:genel} 
\end{eqnarray}
where $m$ is the mass of the scalar particle, $\xi$ is the conformal coupling constant ($\xi = \frac{3}{16}$ in five dimensions) $R$ is the five-dimensional curvature scalar. A dot ($^.$) denotes differentiation w.r.t $\eta$ and a prime ($\prime$) denotes differentiation w.r.t $\sigma$.
For simplicity we choose to investigate a massless field ($m = 0$). 
To quantize the field, we impose the following commutation relations between the field and its conjugate momenta as usual
\be
\le[\psi(x^A),\psi(x'^A)\ri] = 0, \le[\pi(x^A),\pi(x'^A)\ri] = 0, \le[\psi(x^A),\pi(x'^A)\ri] = i \delta^3({\vec x}-{\vec x'})\delta(\s-\s').
\ee
We introduce the creation and annihilation operators using the following mode decomposition as
\be
\psi = \f{1}{4\pi^2}\int d^4q \le[a_q \Phi_q(x^A) + a_q^\da \Phi_q^*(x^A)\ri], \mbox{where}~~~~ q \equiv \{\vec{k},k_\s\}
\ee
where $\Phi_q(x^A)$ are the eigen-solutions of the field equation and $d^4q/4\pi^2$ is the integral measure in five dimensional momentum space.
This decomposition implies
\be
[a_q,a_q'] = [a_q^\da, a_{q'}^\da] = 0, [a_q,a_{q'}^\da] = \delta^4(q-q').
\ee
The vacuum state is defined as $a_q|0\> =0$ for all $q$.

To separate variables in the field equation we use the following ansatz:  
\begin{equation}
 \Phi_q (\eta,{\bf x},\sigma) = \frac{1}{N} \phi_q (\eta)F({\bf x})G(\sigma) .
 \label{eq:separating}
\end{equation}
where $N = e^fa\sq{b}$.
%Putting Eq.(\ref{eq:separating}) in Eq.(\ref{eq:genel}), we get,
%\newpage
%\begin{eqnarray}
%& &\left [\frac{\ddot \phi_q(t)}{\phi_q(t)} - \frac{3\ddot a}{8a} - 
%\frac{\ddot% \phi}{8\phi} - \frac{3\dot a^2}{8a^2} + \frac{\dot \phi^2}{4\phi^2}% - \frac{3\dot a\dot \phi}{8a\phi} \right ] - \frac{1}{a^2} \left[ \frac{1}
%{F({\bf x})}\left\{ \frac {d^2F({\bf x})}{dx^2} +\frac {d^2F({\bf x})}{dy^2} +
% \frac {d^2F({\bf x})}{dz^2} \right\} \right] \nonumber \\ & & - \frac{1}{\phi
%^2} \left[ e^{2f} \left \{\frac {G''(\sigma)}{G(\sigma)} + 2 f' \frac{G'(\sigma%)}{G(\sigma)} + \left(\frac{f''}{2} + \frac{3f'^2}{4}\right) \right\}\right] = %0 .\label{eq:aftersep}
%\end{eqnarray}
The normalization condition for $\psi$ or $\Phi_q$ gives the Wronskian relation,
\begin{equation}
\dot \phi_q^*\phi_q - \dot\phi_q\phi_q^* = i  .\label{eq:wronskian}
\end{equation}
Let us set,
\begin{equation}
\frac{1}{F({\bf x})}\left\{ \frac {d^2F({\bf x}) }{dx^2}+ \frac {d^2F({\bf x})}{dy^2} + \frac {d^2F({\bf x})}{dz^2} \right\} = -{\bf k}^2 ,\label{eq:Fx}
\end{equation}
and
\begin{equation}
e^{2f} \left\{\frac {G''(\sigma)}{G(\sigma)} + 2 f'\frac{G'(\sigma)}{G(\sigma)} + \left(\frac{f''}{2} + \frac{3f'^2}{4}\right) \right\} = - k_\sigma^2 .
\label{eq:Gsigma}
\end{equation}

The above two considerations imply,
%\begin{equation}
%\ddot \phi_q(t) + \left[ \left( \frac{\bf k^2}{a^2} + \frac{k_\sigma^2}{ \phi^2%}\right)- \frac{3\ddot a}{8a} - \frac{\ddot \phi}{8\phi} + \frac{3\dot a^2}{8a^%2} + \frac{\dot \phi^2}{4\phi^2} - \frac{3\dot a\dot \phi}{8a\phi} \right]\phi_%l(t) = 0 .
%\label{eq:genchieqn}
%\end{equation}
%By looking at the above equation one can see that this equation remains 
%qualitatively unchanged if we exchange the roles of $a(t)$ and $\phi(t)$.
%
%With respect to  conformal time($\eta$) $\left( \mbox{where,}\ \eta = \int \fra%c{dt}{a(t)} \right)$ the equation corresponding to Eq. (\ref{eq:genchieqn}) com%es out as,
\begin{equation}
\ddot \phi_q(\eta) + \left[ \left({\bf k}^2 + \frac{a^2}{b^2} k_\sigma^2 \right) + \frac{\ddot a}{8a} - \frac{\ddot b}{8b} + \frac{\dot b^2}{4b^2} - \frac{\dot a\dot b}{4a b} \right]\phi_q(\eta) = 0 .
\label{eq:Phit}
\end{equation}

%For $\phi = \phi_0$, a constant, Eq. (\ref{eq:genchieqn}) becomes,
%\begin{equation}
%\ddot \phi_k(t) + \left[  \left( \frac{{\bf k}^2}{a^2} + \frac{k_\sigma^2}{\phi_0^2}\right) + \frac{3}{8}\left(\frac{\dot a^2}{a^2} - \frac{\ddot a}{a}\right)\right]\phi_k(t) = 0
%\label{eq:genphi0}
%\end{equation}

One can write Eq.(\ref{eq:Phit}). as,
\begin{equation}
\ddot \phi_q(\eta) + \left[ \Omega_q^2(\eta) + Q(\eta) \right]\phi_q(\eta) = 0 ,\label{eq:Phit1}
\end{equation}
where,
\begin{equation}
\Omega_q^2(\eta) = \left({\bf k}^2 + \frac{a^2}{b^2} k_\sigma^2 \right)\label{eq:Omgforms}
\end{equation}
and
\begin{equation}
Q(\eta) =\frac{\ddot a}{8a} - \frac{\ddot b}{8b} + \frac{\dot b^2}{4b^2} - \frac{\dot a\dot b}{4a b}.\label{eq:Qforms}
\end{equation}
Note that for an observer on the brane this scalar field (with a specific $k_\s$), which is massless in the bulk, appears to be a scalar field with mass `$k_\s$' \cite{Ghosh:2008zs} with an effective on brane scale factor $a(\eta)/b(\eta)$. Thus different $k_\s$ modes, if excited, may be interpreted as particles with different masses on the brane as such.
Moving ahead, we note that, Eq. \ref{eq:Phit1} admits WKB solutions of the form \cite{parker:2012,Zeldovich:1971mw},
\begin{equation}
\phi_q(\eta) = \frac{\alpha_q(\eta)}{\sqrt{2\Omega_q}} e_-  + \frac{\beta_q(\eta)}{\sqrt{2\Omega_q}} e_+ , ~~~~\mbox{where}~~~~ e_{\pm} = e^{\pm i\int^\eta \Omega_q d\eta}
\label{eq:WKBsol}
\end{equation}
with a further restriction,
\begin{equation}
\dot\phi_q(\eta) = -i\Omega_q \left[\frac{\alpha_q(\eta)}{\sqrt{2\Omega_q}} e_- - \frac{\beta_q(\eta)}{\sqrt{2\Omega_q}} e_+\right] ,\label{eq:WKBrestriction}
\end{equation}
where $\alpha_q$ and $\beta_q$ are the Bogoliubov coefficients.

Putting Eqs.(\ref{eq:WKBsol}) and (\ref{eq:WKBrestriction}) in Eqs.(\ref{eq:Phit1}) and using the condition  (\ref{eq:wronskian}) we get,
\begin{eqnarray}
\dot \alpha_q& = &\frac{1}{2} \left( \frac{\dot \Omega_q}{ \Omega_q} - i\frac{Q}{\Omega_q}\right)\beta_q\ e_+ - i \frac{Q}{2 \Omega_q}\alpha_q  ,\nonumber\\
                                             \label {eq:alphabetadot}       \\
\dot \beta_q& = &\frac{1}{2} \left( \frac{\dot \Omega_q}{ \Omega_q} + i\frac{Q}{\Omega_q}\right)\alpha_q\ e_- + i \frac{Q}{2 \Omega_q}\beta_q ,\nonumber
\end{eqnarray}
\begin{equation}
\mbox{and,} \hspace{2cm} |\alpha_q|^2 - |\beta_q|^2 = 1 .\label{eq:modsqrdiff}
\end{equation}

Corresponding initial conditions are,
\begin{equation}
 \alpha_q (\eta_0) = 1 \hspace{0.5cm} \mbox{and}\hspace{0.5cm} \beta_q(\eta_0) = 0. \label{eq:initial1}
\end{equation}
where, $\eta_0$ is a suitably chosen initial time of a chosen adiabatic vacuum\footnote{e.g. one may choose $a(\eta_0) = b(\eta_0) = 1$.}. 
Then the number of particles thus created in mode $q$ is given by,
\begin{equation}
N_q = \lim_{\eta\rightarrow \infty} |\beta_q|^2 .\label{eq:partno}
\end{equation}

%But $\beta_q$ normally depends on $t_0$ (which is nearly of the order of Planck time) in a complicated manner. So it is convenient to address the problem in conformal time coordinate to get some exact numerical values of particle number density or energy density.

Following Zel'dovich and Starobinsky \cite{Zeldovich:1971mw}, Eqs.(\ref{eq:alphabetadot}) and (\ref{eq:modsqrdiff}), with the initial conditions (\ref{eq:initial1}), can be recast in the form,
\begin{eqnarray}
\dot s_q & = & \frac{\dot \Omega_q}{2 \Omega_q}u_q + \frac{Q}{2 \Omega_q}\t_q \la{eq:s}, \\
\dot u_q & = & \frac{\dot \Omega_q}{\Omega_q}(1 + 2s_q) - \left[\frac{Q}{\Omega_q} + 2 \Omega_q \right]\t_q \la{eq:u},\\ 
\dot \t_q & = & \frac{Q}{\Omega_q}(1 + 2s_q) + \left[\frac{Q}{\Omega_q} + 2 \Omega_q \right]u_q \la{eq:t},
\end{eqnarray}
where,
\ba
s_q = |\beta_q|^2, ~~  u_q = 2 Re\left[ \alpha_q \beta_q^* e_-^2\right], ~~ \t_q = 2i Im\left[ \alpha_q \beta_q^* e_-^2\right] \la{eq:sut-def}.%\nonumber
\ea
are the ZS variables which obey the following initial conditions,
\begin{equation}
s_q(\eta_0) = u_q(\eta_0) = \t_q(\eta_0) = 0. \label{eq:sut-conds}
\end{equation}
Thus $s_q(\eta)$ represents the particle number density and is related to the energy density in the $\eta \ra \infty$ or the adiabatic limit. Similarly, $u_q(\eta)$ and $\t_q(\eta)$ are essentially related to the pressure and current density components of the EMT of the created particles\footnote{Note that, in case of EMT of a spin-$\f{1}{2}$ field, corresponding axial anomaly or the axial current in $m\ra 0$ limit is equal to $\t_q$ integrated over all the modes.}. 

To get the number of particles created in mode $q$ one has to solve the first order system of  differential equations (\ref{eq:s}-\ref{eq:t}) with initial conditions (\ref{eq:sut-conds}) and determine $s_q$ when $\eta\rightarrow \infty $. These equations can be evolved numerically using standard codes, in cases where one is unable to find an analytic solution \cite{Garriga:1989jx,Ghosh:2016epo}.
In the following, we solve Eqs. \ref{eq:s}-\ref{eq:t} analytically using the adiabatic approximation. In the adiabatic or quasi classical regime $|\dot\O_q|<<|\O_q|^2$. This means if we expand the ZS variables in adiabatic series as 
\be
s_q = \sum_{r=1}^\infty s_q^{(r)}, u_q = \sum_{r=1}^\infty u_q^{(r)}, \t_q = \sum_{r=1}^\infty \t_q^{(r)}, \la{eq:sut-series}
\ee
where, `$r$' denotes the adiabatic order\footnote{An adiabatic order essentially equals the number of time derivative of $a(\eta)$ contained in the term.}, in the adiabatic limit e.g. $s_q^{(r)}$ is suppressed as $|\dot\O_q / \O_q^2|^r$ and $s_q^{(r+1)}/s_q^{(r)} << 1$. 
If we replace Eq. \ref{eq:sut-series} in Eqs. \ref{eq:s}, \ref{eq:u} and \ref{eq:t}, we get the following recursion relations among the terms of different adiabatic orders of $s_q$, $u_q$ and $\t_q$,
\ba
\mbox{Eq.} \ref{eq:s} \implies ~~ s_q^{(r)} &=& \int d\eta \le(\f{\dot\O_q}{2\O_q} u_q^{(r)} + \f{Q}{2\O_q}\t_q^{(r-1)}\ri), \la{eq:s-ad}\\
\mbox{Eq.} \ref{eq:u} \implies ~~ \t_q^{(r+1)} &=&  \f{\dot\O_q}{\O_q^2} s_q^{(r)} - \f{1}{2\O_q} \dot u_q^{(r)} - \f{Q}{2\O_q^2}\t_q^{(r-1)},  \la{eq:u-ad} \\
\mbox{Eq.} \ref{eq:t} \implies ~~ u_q^{(r+2)} &=& \f{\dot\t_q^{(r+1)}}{2\O_q} - \f{Q}{2\O_q^2}(2s_q^{(r)}+u_q^{(r)}), \la{eq:t-ad}
\ea
where $r = 2,4,..$ and 
\be
\t_q^{(1)} = \f{\dot\O_q}{2\O_q^2}, ~~ s_q^{(1)}=u_q^{(1)}=\t_q^{(2)}=0 \implies s_q^{(r)}=u_q^{(r)}=\t_q^{(r+1)}=0 ~~~~\mbox{for odd}~r.
\ee
Using these results back in Eqs \ref{eq:s-ad}, \ref{eq:u-ad} and \ref{eq:t-ad}, we get, 
\be
u_q^{(2)} = \f{\dot\t_q^{(1)}}{2\O_q} - \f{Q}{2\O_q^2} = \f{1}{4\O_q^2} \le(-2\f{\dot\O_q^2}{\O_q^2} + \f{\ddot\O_q}{\O_q} - 2Q\ri) \implies s_q^{(2)} = \f{\dot\O_q^2}{16\O_q^4}.
\ee
The higher order terms can be derived in a straightforward manner, upto arbitrary order, from Eqs \ref{eq:s-ad}, \ref{eq:u-ad} and \ref{eq:t-ad} (see Appendix \ref{app:sut}).

\section{Energy momentum tensor}

The energy momentum tensor of a conformally coupled field is given by \cite{Zeldovich:1971mw,Flanagan:1996gw}
\be
T_{AB} = \pa_A\psi^* \pa_B\psi - \f{g_{AB}}{2}\le[|\na\psi|^2 + m^2|\psi|^2\ri] + \xi \le(G_{AB} + g_{AB}\square - \na_A\na_B\ri)|\psi|^2
\ee
where $G_{AB}$ is the Einstein tensor and $\square = g^{AB}\na_A\na_B$. For metric (\ref{eq:metric}) with separation of variables given by Eq. (\ref{eq:separating}), the vacuum expectation value of the 00-component of the stress tensor (with respect to the adiabatic vacuum ) i.e. the energy density created during the evolution of warped cosmological braneworlds is given by,
\ba
\< T_{00} \> = \int \f{d^4q}{(2\pi N)^2} && \le[|G|^2\le\{\f{|\dot\phi_q|^2}{2} + \f{1}{16}\le(\f{\dot a}{a} - \f{\dot b}{b}\ri) \pa_\eta |\phi_q|^2 + \le(\f{k^2}{2} - \f{1}{16}\le(\f{\dot a}{a} - \f{\dot b}{b}\ri)^2 \ri)|\phi_q|^2\ri\} \ri. \nn \\ 
&& \le. + \f{a^2e^{2f}}{2b^2} |\phi_q|^2 \le\{|G'|^2 - \f{5f'}{8}\pa_\s |G|^2 - \le(\f{f'^2}{2}+ \f{3f''}{8}\ri)|G|^2 - \f{3}{8} \pa_\s^2 |G|^2\ri\} \ri]  \\
 = \int \f{d^4q}{(2\pi N)^2} && |G|^2 \le[\f{|\dot\phi_q|^2}{2} + \f{\g}{16} \pa_\eta |\phi_q|^2 + \le(\f{{}_0\O_q^2}{2} - \f{\g^2}{16} \ri)|\phi_q|^2 \ri]    \la{eq:T00}
\ea

where we have defined 
\be
 \g = \le(\f{\dot a}{a} - \f{\dot b}{b}\ri),~~ {}_0\O_q^2(\s) = k^2 + \f{a^2}{b^2}   K_0^2(k_\s;\s),
\ee
where
\be
K_0^2(k_\s;\s) = \f{e^{2f}}{|G|^2} \le\{|G'|^2 - \f{5f'}{8}\pa_\s |G|^2 - \le(\f{4f'^2 + 3f''}{8}\ri)|G|^2 - \f{3}{8} \pa_\s^2 |G|^2\ri\}
\ee

%What remains to be done is to solve the above iterative equations upto the $6th$ adiabatic order and write down the components of $T_{AB}$ in terms of $s_q$, $u_q$ and $\t_q$. Then one can proceed with the regularization of the energy momentum tensor by adiabatic subtraction of the leading order terms which are responsible for divergence in energy density as such. 

Using equations \ref{eq:WKBsol}, \ref{eq:WKBrestriction} and \ref{eq:sut-def}, Eq. \ref{eq:T00} can be written as
\ba
\< T_{00} \> = \int \f{d^4q}{(2\pi N)^2} |G|^2 && \le[ \f{\O_q}{2}\le(\f{1}{2} + s_q - u_q\ri) + \f{{}_0\O_q^2}{2\O_q} \le(\f{1}{2} + s_q + u_q\ri) \ri. \nn \\
&& ~~~~ \le. - \f{\g}{16} \t_q  - \f{\g^2}{16\O_q} \le(\f{1}{2} + s_q + \f{u_q}{2}\ri)\ri]   \la{eq:T00_sut}
\ea
Note that the $1/2$-factors appearing in Eq. \ref{eq:T00_sut} (and in the following equations) represent the vacuum contributions (which will later be subtracted for regularization). The above expression for energy density does contain features of a dynamic warped extra dimension. In the limit $a(\eta)=b(\eta)$ and $f(\s)=0$, we have $\g=0$ and $\O_q^2 = {}_0\O_q^2 = k^2 + k_\s^2$, which would make all the terms in Eq. \ref{eq:T00_sut} vanish to recover the standard result that no particles would be created from a massless scalar field in a conformally flat spacetime.

Similarly, the other components of the energy momentum tensor can be written, e.g. the pressure density along the on-brane spatial directions, as
\ba
\< T_{ii} \> = \int \f{d^4q}{(2\pi N)^2} && |G|^2\le[\f{|\dot\phi_q|^2}{2} - \f{\g}{16} \pa_\eta |\phi_q|^2  - \f{3}{16} \pa^2_\eta |\phi_q|^2  + \le(k_i^2 - \f{{}_{i}\O_q^2}{2} - \f{\g^2}{16} \ri)|\phi_q|^2 \ri]  \la{eq:Tii} \\ 
%= \int \f{d^4q}{N^2} && |G|^2 \le[\le(\f{k_i^2}{\O_q} - \f{\O_q^2 - {}_i\O_q^2}{2\O_q} \ri) \le(\f{1}{2} + s_q\ri) +    \le(\f{k_i^2}{2\O_q} + \f{\O_q^2 - 2{}_i\O_q^2}{8\O_q}\ri)  u_q \ri. \nn \\
%&& \le. ~~~~~~~~  - \f{\g^2 -6Q}{16\O_q} \le(\f{1}{2} + s_q + \f{u_q}{2}\ri) + \f{\g}{16} \t_q \ri]   \\
= \int \f{d^4q}{(2\pi N)^2} && |G|^2 \le[ \f{k_i^2}{\O_q}  \le(\f{1}{2} + s_q + \f{u_q}{2} \ri) -    \f{\O_q}{2} \le(\f{1}{2} + s_q - \f{u_q}{4} \ri) + \f{{}_i\O_q^2}{2\O_q}\le(\f{1}{2} + s_q - \f{u_q}{2} \ri) \ri. \nn \\
&& \le. ~~~~~~~~  - \f{\g^2 -6Q}{16\O_q} \le(\f{1}{2} + s_q + \f{u_q}{2}\ri) + \f{\g}{16} \t_q \ri]    \la{eq:Tii_sut}
\ea

where,
\be
{}_i\O_q^2(\s) = k^2 + \f{a^2}{b^2}  K_i^2(k_\s;\s)
\ee
and 
\be
K_i^2(k_\s;\s) = \f{e^{2f}}{|G|^2} \le\{|G'|^2 + \f{3f'}{8}\pa_\s |G|^2 + \le(\f{5f'^2 - 3f''}{8}\ri)|G|^2 - \f{3}{8} \pa_\s^2 |G|^2\ri\}
\ee

The pressure density along the extra spatial direction `$\s$' is given by
\ba
\< T_{\s\s} \> = \int \f{d^4q}{(2\pi N)^2} && \f{b^2|G|^2}{a^2e^{2f}}\le[\f{|\dot\phi_q|^2}{2} - \f{\g}{8} \pa_\eta |\phi_q|^2  - \f{3}{16} \pa^2_\eta |\phi_q|^2  + \le(\f{a^2}{b^2} k_\s^2 - \f{{}_{\s} \O_q^2}{2} + \f{\g^2}{8} - \f{3Q}{2} \ri)|\phi_q|^2 \ri]  \la{eq:Tss}   \\ 
%= \int \f{d^4q}{N^2} && \f{b^2|G|^2}{a^2e^{2f}} \le[\le(\f{a^2}{b^2}\f{k_\s^2}{\O_q} + \f{\O_q^2 - {}_\s\O_q^2}{2\O_q} \ri) \le(\f{1}{2} + s_q\ri) +    \le(\f{a^2}{b^2}\f{k_\s^2}{2\O_q} + \f{\O_q^2 - 2{}_\s\O_q^2}{8\O_q}\ri)  u_q \ri. \nn \\
%&& \le. ~~~~~~~~~~  + \f{\g^2 - 9Q}{8\O_q} \le(\f{1}{2} + s_q + \f{u_q}{2}\ri) + \f{\g}{8} \t_q \ri]    \\
= \int \f{d^4q}{(2\pi N)^2} && \f{b^2|G|^2}{a^2e^{2f}} \le[\le(\f{a^2}{b^2}\f{k_\s^2}{\O_q} + \f{\O_q}{2}\ri) \le(\f{1}{2} + s_q + \f{u_q}{4}\ri) -  \f{{}_\s\O_q^2}{2\O_q} \le(\f{1}{2} + s_q + \f{u_q}{2}\ri) \ri. \nn \\
&& \le. ~~~~~~~~~~  + \f{\g}{8} \t_q + \f{\g^2 - 9Q}{8\O_q} \le(\f{1}{2} + s_q + \f{u_q}{2}\ri) \ri]    \la{eq:Tss_sut}
\ea

where,
\be
{}_\s\O_q^2(\s) = k^2 + \f{a^2}{b^2}  K_\s^2(k_\s;\s)
\ee
and 
\be
K_\s^2(k_\s;\s) = \f{e^{2f}}{|G|^2} \le\{|G'|^2 + \f{f'}{2}\pa_\s |G|^2 +  \f{7f'^2}{8}|G|^2\ri\}.
\ee

The only off-diagonal element of EMT, which is a distinguishing feature of such warped spacetimes is given by
\ba
\< T_{0\s} \> &=& \int \f{d^4q}{(2\pi N)^2} |G|^2 \le[3 \f{\dot{b}}{b}f' |\phi_q|^2 - K_{0\s}^2 \le( \pa_\eta |\phi_q|^2 - 2 \f{\dot\a}{\a}|\phi_q|^2\ri) \ri]  \la{eq:T0s}   \\ 
&=& \int \f{d^4q}{(2\pi N)^2} |G|^2  \le[3 \f{\dot{b}}{b\O}f' \le(\f{1}{2} + s_q + \f{u_q}{2}\ri) + K_{0\s}^2 \le\{\t + 2 \f{\dot\a}{\a}\le(\f{1}{2} + s_q + \f{u_q}{2}\ri)\ri\} \ri] \la{eq:T0s_sut}
\ea
where
\be
K_{0\s}^2(k_\s;\s) = \f{1}{|G|^2} \le(\pa_\s |G|^2 - 3f'|G|^2\ri)
\ee

After getting the expression of bulk EMT we note that these components of bulk EMT are functions of both time and the extra dimension as expected. In the following we explore what these dependencies physically mean.

\subsection{Localisation of matter density}

The functional dependence of the EMT components on time and extra dimension can be studied as a two dimensional surface in $T,t,\s$ space. So, these components can be analyzed graphically. However, these surfaces, in general, will have complicated shapes. Further, it would be difficult to carry out the integration over $k_\s$ to find how $\< T_{AB}\> $ (or its renormalized version) varies along $\s$. Note that, the $\s$-dependence in $\< T_{00}\> $, for example, comes from an over all factor of $|G|/e^f$ and ${}_0\O_q(\s)$ (which carries $K_0^2(k_\s,\s)$).
%In general, $\s$-dependence in $\< T_{AB}\> $ comes from the factors $K_A^2(k_\s;\s)$ (for $A = t,i,\s $) except the overall factor. 
For simplicity, let us look how these factors depend on $\s$, for few fixed values of $k_\s$. This will let us compare, e.g. how much matter is created at different location along the extra dimension due to the bulk scalar field. We can do this even before the adiabatic subtraction as regularization will only effect Z-S variables which do not depend on $\s$. 

Firstly, we look at what happens to the factors $K_A^2(k_\s,\s)$ at the location of the brane. 
Eq. \ref{eq:Gsigma}, near $\s=0$, in presence of a growing warp factor, reduces to
\be
G''(\s) + \le(\f{\k^2}{2} + k_{\s}^2\ri)G(\s) = 0, \la{eq:Gnear0}
\ee
i.e. $G(\s)$ behaves like a plane wave around $\s=0$. Eq. \ref{eq:Gnear0} further implies that, near $\s=0$
\be
K_0^2 = K_i^2 = k_{\s}^2 + \f{\k^2}{8} , ~~~~ K_\s^2 = k_{\s}^2 + \f{\k^2}{2}.
\ee
Thus near $\s=0$, ${}_A\O_q(\s)$ may be considered as independent of $\s$. Similar $\s$-dependence (or independence) is found for $\< T_{ii}\> $ as the pre-factor is same. In $\< T_{\s\s}\> $, there is an extra factor of $e^{-2f}$ present which does not change any qualitative behavior of the pre-factor in any significant manner near $\s=0$. This simple analysis tells us it is enough to analyze the pre-factor in order to understand the spatial variation of EMT components. 

In the following, we have graphically presented the behaviour of the pre-factor in the integrand of $\< T_{00}\> $ i.e. $|G|^2e^{-2f}$ as a function of $\s$ (for suitable lowest values of $k_\s$). In order to do that, we have solved Eq. \ref{eq:Gsigma} numerically (with initial conditions $G(0)=1$ and $G'(0)=0.1$) for two different functional forms of the warp factor, namely $f(\s)=\pm \log(\cosh \k\s)$, where $1/\k$ represents a length scale along the extra dimension and the `+' sign (`-' sign) represents the so-called growing (decaying) warp factor in thick brane models. We have set $\k=1$ for all numerical computations.

\begin{figure}[!ht]
\includegraphics[width=2.5in]{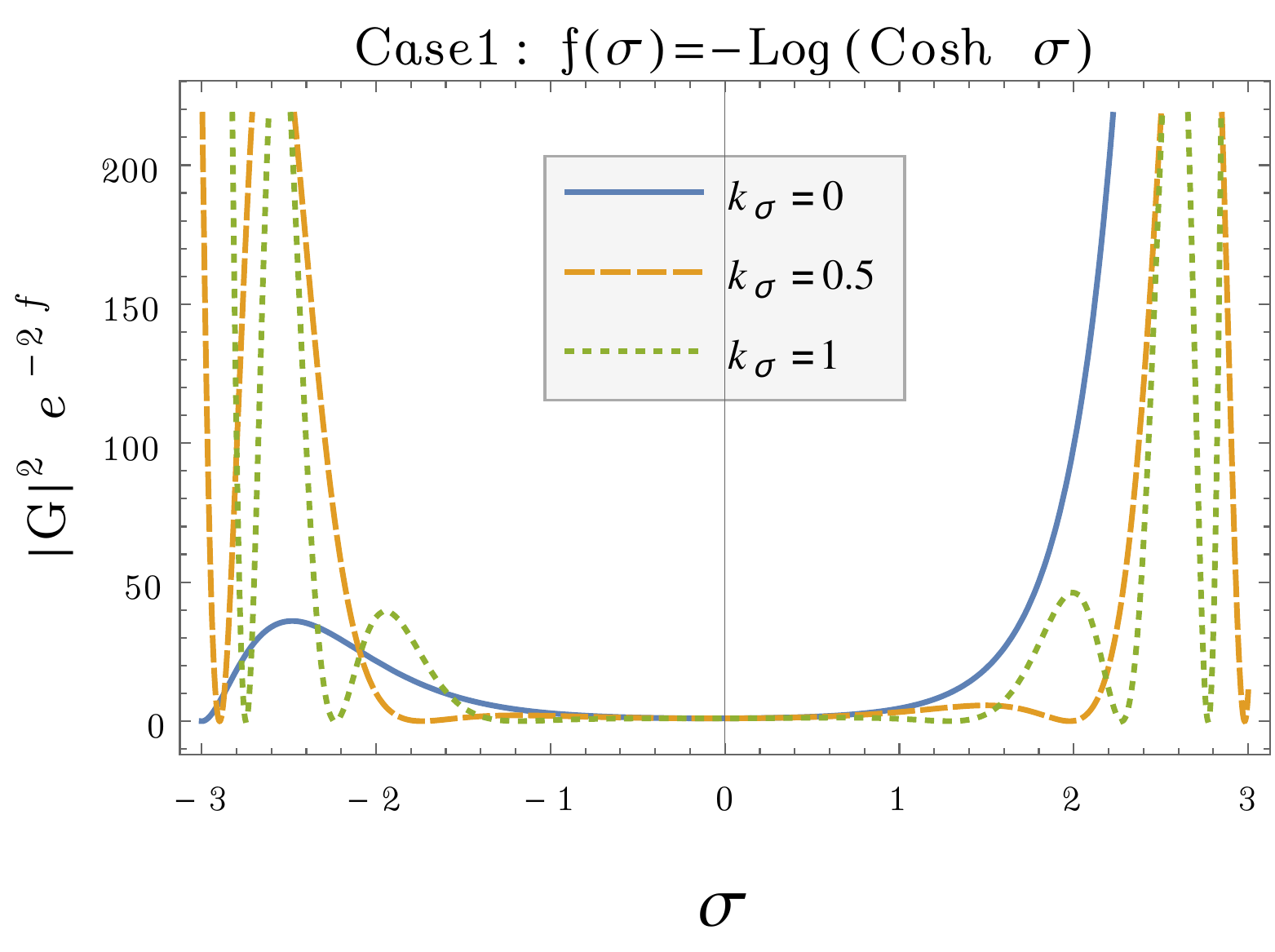} \hspace{1cm}
\includegraphics[width=2.5in]{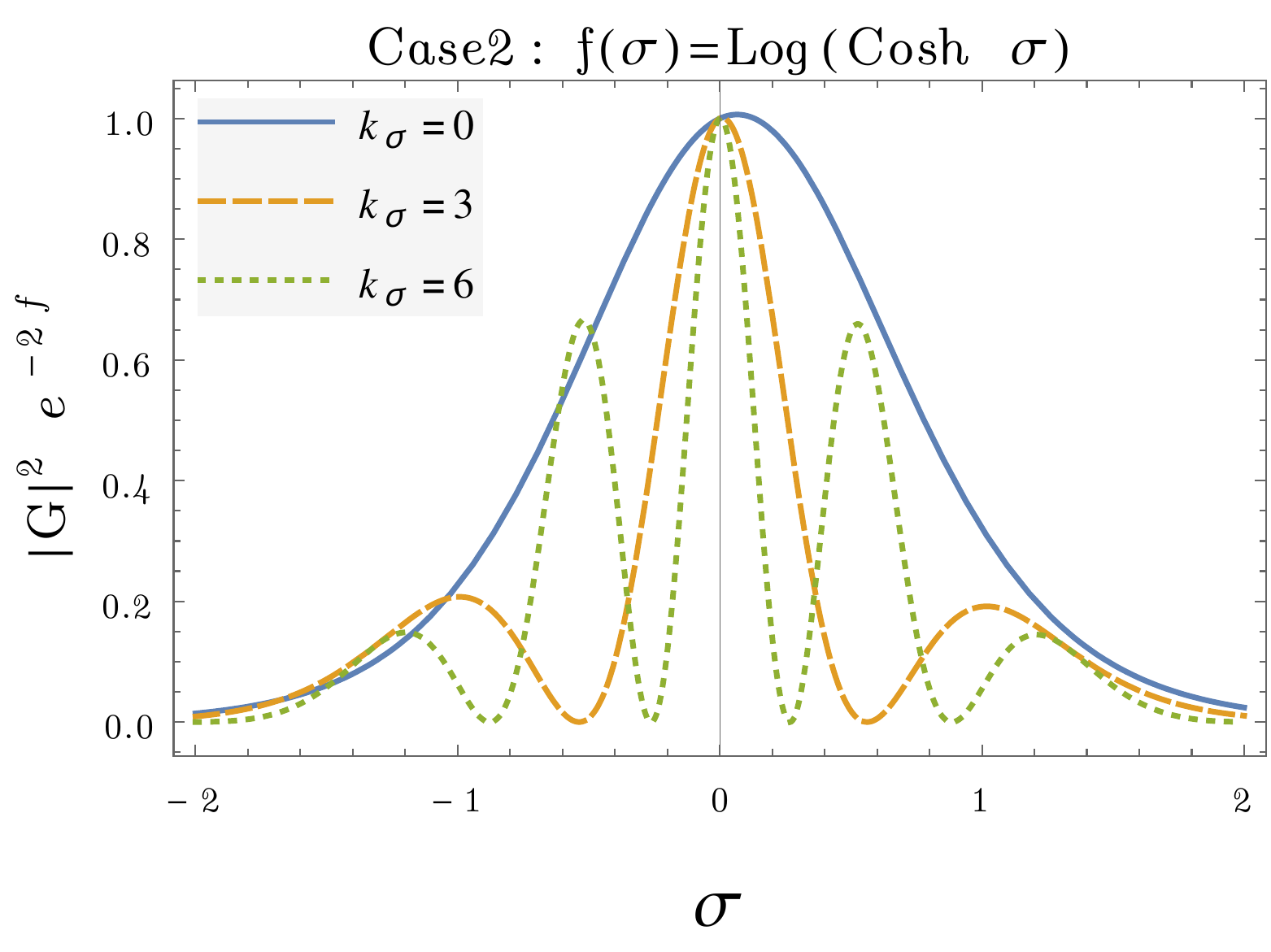} %[width=2.2in,angle=-90] 
\caption{Left: Variation of the overall pre-factor in presence of decaying warp factor for $k_\s = 0, 0.5, 1$. Right: Variation of the overall pre-factor in presence of growing warp factor for $k_\s = 0, 10, 20$.} \label{fig:case1and2} 
\end{figure}

\begin{figure}[!ht]
\includegraphics[width=2.5in]{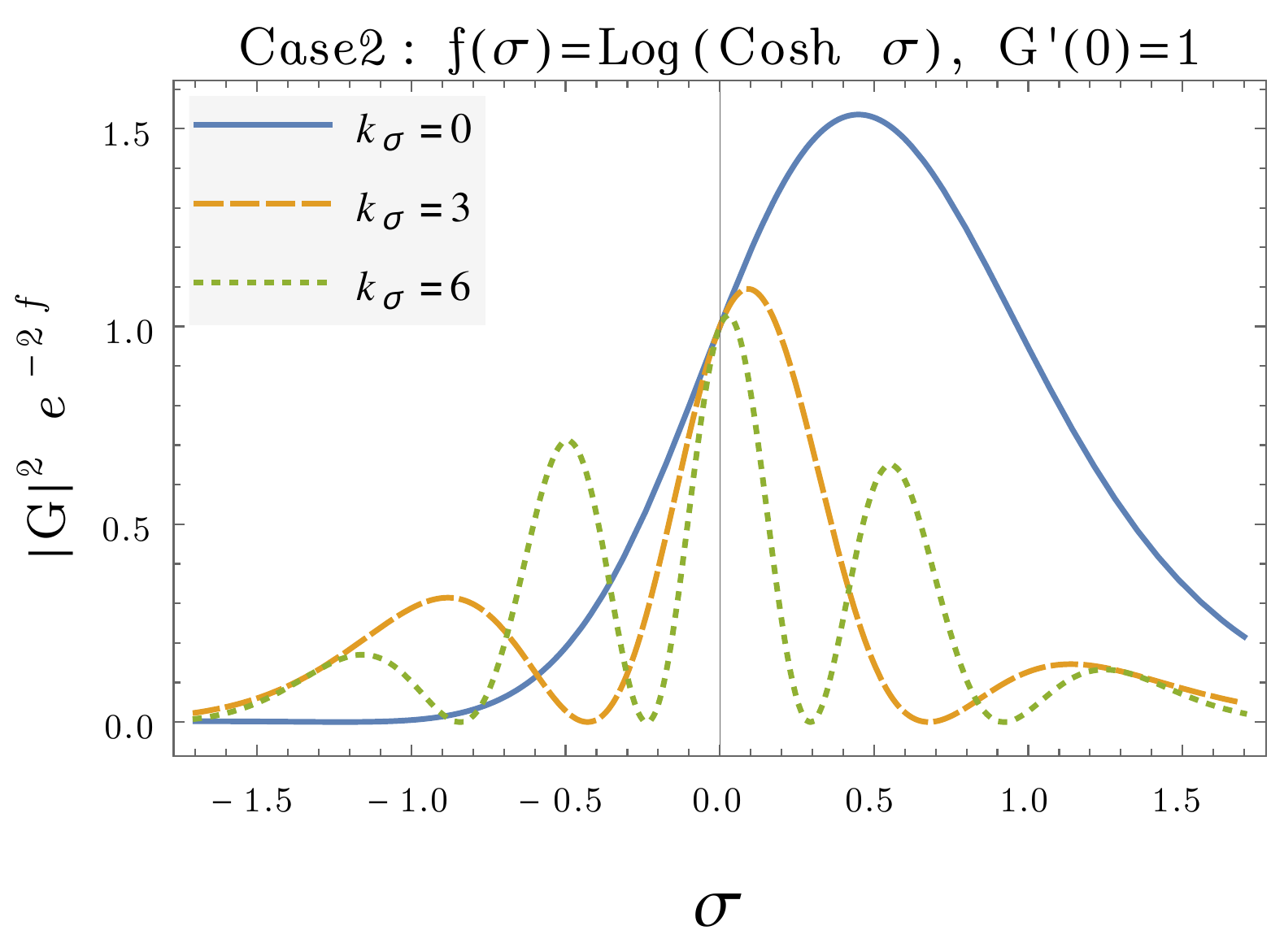} \hspace{1cm}
\includegraphics[width=2.5in]{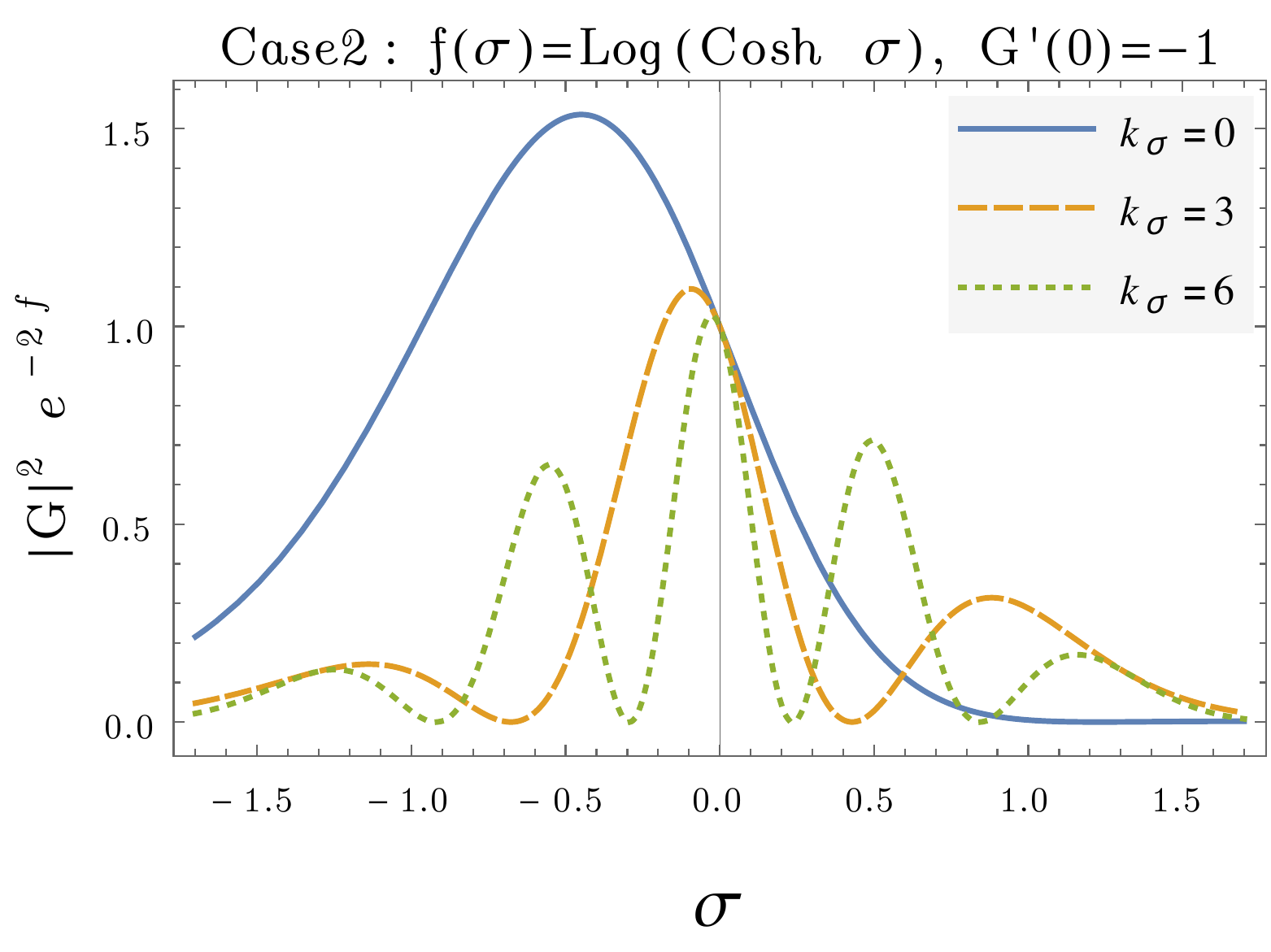} %[width=2.2in,angle=-90] 
\caption{Left: The overall pre-factor in presence of growing warp factor for $G(0)=1$ and $G'(0)=1$. Right: The overall pre-factor in presence of growing warp factor for $G(0)=1$ and $G'(0)=-1$.} \label{fig:case21and22} 
\end{figure}

The numerical values of $k_\s$ are chosen\footnote{Allowed values of $k_\s$ are discrete in case of finite extra dimension \cite{Ghosh:2008zs}.} such that the curves show one, three and five minima or maxima in Fig. \ref{fig:case1and2}. The left graph with the decaying warp factor implies that most of the energy density is created away from the brane location ($\s=0$). Further, density increases indefinitely in the $\s \ra \pm \infty$ limit. Thus this scenario is not physically interesting as it  may only be interpreted as creation of two {\em physical} branes, formed due to accumulation of infinite amount of matter, separated by infinite distance. On the other hand, the right graph with the growing warp factor implies that most of the created energy density is localised near $\s=0$. Thus a thick brane is formed with localised matter with masses $k_\s$. Thickness of this brane may be quantified by the `width' of the envelope of all the pre-factor curves for all $k_\s$ modes (the zero-mode curve itself looks like such an envelope!). This scenario is physically more appealing. Therefore we carry on our analysis on braneworlds with growing warp factor only. Note that the central maxima for non-zero $k_\s$ modes lie on or near $\s=0$, however, the zero mode seems to be little off. The location of this peak depends on the initial conditions used. To verify how sensitive the location of central maxima is to initial conditions, we have retraced these curves with significantly different initial conditions namely with $G'(0)=1$ (left) and $G'(0)=-1$ (right) in Fig. \ref{fig:case21and22}.

Fig. \ref{fig:case21and22} clearly shows that though the zero mode can peak away from the so-called brane location, localisation of the non-zero modes are insensitive to the initial conditions. So we may conclude that localisation of massive particles on the brane is a generic feature of braneworld with a growing warp factor \footnote{Note that, earlier, {\em classical} localisation of massive particles in such spacetimes has been demonstrated through geodesic confinement in \cite{Ghosh:2009ig}}.

\subsection{Adiabatic regularization of bulk EMT}

Let us now turn to the regularization of the components of $\< T_{AB}\> $ which obviously contains ultra-violet divergences, coming from the large modes, apart from the usual Minkowski space divergence. According to the adiabatic subtraction algorithm, the REMT is found after subtracting the vacuum contributions and the terms upto the fourth adiabatic order from the adiabatic series expansion of EMT. Note that as the adiabatic regularization is done mode by mode, components of REMT of different adiabatic order and of each mode totally decouple from each other and are covariantly conserved by definition.

In our case, let us define the on-brane {\em conformal} Hubble parameter to be $h = \dot{a}/{a}$. 
Note that, as long as $a k_{\s}/b << h$, the adiabatic corrections (which go as powers of $|\dot\O_q / \O_q^2|$) are much smaller than the zeroth order ones even for the $k = 0$ mode making the adiabatic expansion legitimate for all $k$-modes.
On the other hand, for $k_{\s}=0$ mode, the adiabatic expansion fails for modes with $k < h$ (i.e. for on-brane superhorizon modes). However, the expansion is still valid for the mode functions satisfying $k >> h$ (i.e.
for on-brane subhorizon modes). Thus the momentum integrals can be performed in the interval ($k_* , \infty$), for some $k_*>>h$. Since the divergences are essentially caused by large modes we expect the adiabatic regularization to be valid in this case. Subtracting terms upto fourth adiabatic order from the diverging energy density, we get the renormalized energy density to be,
\ba
\< T_{00} \> {}_{Ren} 
&=& \int \f{d^4q}{(2\pi N)^2} |G|^2 \le[\le( \f{\O_q^2 + {}_0\O_q^2}{2\O_q} - \f{\g^2}{16\O_q}\ri)\le(s_q - s_q^{(2)} - s_q^{(4)}\ri) \ri. \nn \\ 
~~~~ && \le. - \le(\f{\O_q^2 - {}_0\O_q^2}{2\O_q} + \f{\g^2}{32\O_q} \ri) \le(u_q - u_q^{(2)} - u_q^{(4)}\ri) - \f{\g}{16} \le(\t_q - \t_q^{(1)} - \t_q^{(3)}\ri) \ri]   \la{eq:T00ren} \\
& \approx & - \int \f{d^4q}{(2\pi N)^2} |G|^2 \le[\le( \f{\O_q^2 + {}_0\O_q^2}{2\O_q} - \f{\g^2}{16\O_q}\ri) s_q^{(6)} - \le(\f{\O_q^2 - {}_0\O_q^2}{2\O_q} + \f{\g^2}{32\O_q} \ri) u_q^{(6)} - \f{\g}{16} \t_q^{(5)}\ri] .  \la{eq:T00ren1}
\ea
In the last line we have approximated the renormalized quantity by neglecting terms of order higher than the sixth in the adiabatic series (see Appendix \ref{app:sut}).	Similarly, the other REMT components can be written as

\ba
\< T_{ii} \> {}_{Ren} %= \int \f{d^4q}{N^2} && |G|^2 \le[\le(\f{k_i^2}{\O_q} - \f{\O_q^2 - {}_i\O_q^2}{2\O_q} \ri) \le(\f{1}{2} + s_q\ri) +    \le(\f{k_i^2}{2\O_q} + \f{\O_q^2 - 2{}_i\O_q^2}{8\O_q}\ri)  u_q \ri. \nn \\
%&& \le. ~~~~~~~~  - \f{\g^2 -6Q}{16\O_q} \le(\f{1}{2} + s_q + \f{u_q}{2}\ri) + \f{\g}{16} \t_q \ri]   \\
\approx - \int \f{d^4q}{(2\pi N)^2} && |G|^2 \le[\le(\f{k_i^2}{\O_q} - \f{\O_q^2 - {}_i\O_q^2}{2\O_q} - \f{\g^2 -6Q}{16\O_q} \ri)  s_q^{(6)}  \ri. \nn \\
&& \le. ~~~~~~~~  +    \le(\f{k_i^2}{2\O_q} + \f{\O_q^2 - 2{}_i\O_q^2}{8\O_q}  - \f{\g^2 -6Q}{32\O_q}\ri)  u_q^{(6)} + \f{\g}{16} \t_q^{(5)} \ri]   
  \la{eq:Tiiren}
\ea

\ba
\< T_{\s\s} \> {}_{Ren}
%= \int \f{d^4q}{N^2} && \f{b^2|G|^2}{a^2e^{2f}} \le[\le(\f{a^2}{b^2}\f{k_\s^2}{\O_q} + \f{\O_q^2 - {}_\s\O_q^2}{2\O_q} \ri) \le(\f{1}{2} + s_q\ri) +    \le(\f{a^2}{b^2}\f{k_\s^2}{2\O_q} + \f{\O_q^2 - 2{}_\s\O_q^2}{8\O_q}\ri)  u_q \ri. \nn \\
%&& \le. ~~~~~~~~~~  + \f{\g^2 - 9Q}{8\O_q} \le(\f{1}{2} + s_q + \f{u_q}{2}\ri) + \f{\g}{8} \t_q \ri]    \la{eq:Tss_sut}\\
\approx - \int \f{d^4q}{(2\pi N)^2} && \f{b^2|G|^2}{a^2e^{2f}} \le[\le(\f{a^2}{b^2}\f{k_\s^2}{\O_q} + \f{\O_q^2 - {}_\s\O_q^2}{2\O_q} + \f{\g^2 - 9Q}{8\O_q}\ri)  s_q^{(6)}  \ri. \nn \\
&& \le. ~~~~~~~~~~ +    \le(\f{a^2}{b^2}\f{k_\s^2}{2\O_q} + \f{\O_q^2 - 2{}_\s\O_q^2}{8\O_q} + \f{\g^2 - 9Q}{16\O_q}\ri)  u_q^{(6)}  + \f{\g}{8} \t_q^{(5)} \ri]  \la{eq:Tssren}
\ea

\ba
\< T_{0\s} \> {}_{Ren} %&=& \int \f{d^4q}{N^2} |G|^2  \le[3 \f{\dot{b}}{b\O}f' \le(\f{1}{2} + s_q + \f{u_q}{2}\ri) + K_{0\s}^2 \le\{\t + 2 \f{\dot\a}{\a}\le(\f{1}{2} + s_q + \f{u_q}{2}\ri)\ri\} \ri] \\
&\approx & \int \f{d^4q}{(2\pi N)^2} |G|^2  \le[ \le(3 \f{\dot{b}f'}{b\O} + 2 \f{\dot\a}{\a}K_{0\s}^2  \ri) \le(s_q^{(6)} + \f{u_q^{(6)}}{2}\ri) + K_{0\s}^2 \t^{(5)} \ri] \la{eq:T0sren}
\ea
Using the expressions of $s_q^{(6)}$, $u_q^{(6)}$ and $\t^{(5)}$ as given in Appendix \ref{app:sut}, one gets analytic expressions for all the non-vanishing components of the REMT of a massless bulk quantized scalar field in cosmological braneworld background.

%\subsubsection{On-brane REMT} 

%As mentioned earlier, different $k_\s$ modes of a bulk scalar field appear as on-brane scalar fields with different masses equal to $k_\s/b(\eta)$ as such. To get REMT for such case one just needs to remove the $k_\s$-integration from above.
The projected renormalised energy density on the cosmological brane at $\s=0$ (in case of growing warp factor) is given by
\ba
\< T_{00} \> {}_{Ren}^{(brane)} 
& \approx & - \int \f{d^4q}{4\pi^2a^2b} \le[ \O_q s_q^{(6)} + \f{e^2a^2\k^2}{16b^2\O_q} \le(s_q^{(6)} + u_q^{(6)}\ri)  - \f{\g^2}{32\O_q}\le(s_q^{(6)} - u_q^{(6)}\ri)  - \f{\g}{16} \t_q^{(5)}\ri] .  \la{eq:T00ren1brane}
\ea
Similarly, the renormalized pressure densities are
\ba
\< T_{ii} \> {}_{Ren}^{(brane)}
 & \approx & - \int \f{d^4q}{4\pi^2a^2b} \le[ \f{k_i^2}{\O_q} s_q^{(6)}  - \f{4k_i^2 - \O_q^2}{8\O_q} u_q^{(6)} + \f{e^2a^2\k^2}{16b^2\O_q}\le(s_q^{(6)} - \f{u_q^{(6)}}{2}\ri) \ri. \\ \nn
&& ~~~~~~~~~~~~~~~\le. + \f{\g^2 - 6Q}{32\O_q}\le(s_q^{(6)} + \f{u_q^{(6)}}{2}\ri) - \f{\g}{16} \t_q^{(5)}\ri] .  \la{eq:Tiiren1brane}
\ea
In the conformally flat limit (which implies $k_i=k_\s$) the above expressions leads to usual five dimensional renormalised EMT for massless scalar field. 
To get physically interesting outputs from the above complicated expressions one needs to look at specific examples of viable cosmological braneworld scenarios as such. A detailed study of such various models will be reported elsewhere.

%\section{The zero mode}

\section{Discussion}

In this article we have used the adiabatic regularization method to study the effect of particle creation due to a massless bulk scalar field in a warped cosmological braneworld scenario. Our main goal was to uncover the distinguishing effects of the key components of the geometry of such models, i.e. the warping factor, the cosmological expansion factor and the dynamic scale of the extra dimension. It is found that the warping factor plays key role in deciding how much of the created matter would be localised at different locations along the extra dimension, thus providing us a guiding principle to locate where a four dimensional thick brane may be formed. In our case, it is found that presence of a decaying warp factor is not suitable for such matter brane to form. On the other hand, a growing warp factor helps matter to accumulate near the minimum of the warp factor which may be considered as the four dimensional thick brane we live in. This further suggests that as many parallel branewrolds will be formed as the number of minima in the warp factor.

Adiabatic regularization is performed on the EMT by subtracting diverging terms upto fourth adiabatic order from the formal vacuum EMT. This gives us a finite bulk REMT which carry the effect of dynamic nature of the extra dimension. It would be an interesting exercise to compare the matter density projected on the brane with the background on-brane critical energy density to test the viability of various models. However, it will depend on the exact dynamic nature of $a(\eta)$ and $b(\eta)$. Note that, for brane dynamics the ratio $a(\eta)/b(\eta)$ plays the role of effective cosmological scale factor. A large class of cosmological braneworld solutions with various combinations of $a(\eta)$ and $b(\eta)$ were found in \cite{Ghosh:2008vc}. Based on the formalism developed here, a detailed quantitative study of how adiabatically regularized EMT components depend on these expansion factors and effects the braneworld cosmology will be reported in a separate article.

Recently renormalization of EMT of spin 1/2 field is achieved under adiabatic regularization scheme \cite{Landete:2013axa,Landete:2013lpa,delRio:2014cha,Ghosh:2015mva,Ghosh:2016epo}. Few studies on fermionic fields in non-factorisable geometries, e.g. in Randall-Sundrum braneworld background has been reported in literature \cite{Ichinose:2002kg,Grossman:1999ra,Koley:2004at}. %However, in most such cases, the background geometries were assumed to be static. 
It will be interesting to carry out an analysis on quantum spinor fields in warped cosmological background, as we have done for a bulk scalar field in this article. We hope to report on that in a future communication.

\section*{Acknowledgements}

This research is supported by a start-up grant awarded by University Grant Commission, India, with grant number no. F.30-420/2018(BSR). Author is thankful to Prof. Sayan Kar for insightful discussions and the `Visitors Program' of the Centre for Theoretical Studies at the Indian Institute of Technology Kharagpur where the initial part of this work was done. 

\appendix

\section{Useful geometric quantities} \la{ap:geo}

%Following are few useful formulas for cosmological braneworld geometry.

Affine connections:
\be
\Gamma^0_{00} = \Gamma^0_{ii} = \Gamma^i_{i0} = \f{\dot{a}}{a}, ~~ \Gamma^i_{i\s} = \Gamma^\s_{0\s} = f'; ~~~~ i=1,2,3.
\ee
%%%%%%%%%%%%%%%
\be
\Gamma^\s_{00} = - \Gamma^\s_{ii} = \f{a^2 e^{2f}}{b^2}f' ,~~~~ \Gamma^0_{\s\s} = \f{b \dot{b}}{a^2 e^{2f}}
\ee

%%%%%%%%%%%%%%%
\be
\mbox{Ricci scalar:}~~ R = \f{e^{2f}}{a^2} \le(6\f{\ddot{a}}{a} + 4\f{\dot{a}\dot{b}}{ab} + 2 \f{\ddot{b}}{b}\ri) - \f{8f'' + 10 f'^2}{b^2}, \la{eq:R}
\ee

%%%%%%%%%%%%%%%%%%%%%
%\ba
%R_{ABCD}R^{ABCD} &=& \f{e^{-2f}}{a^4}  \la{eq:Rmn-sq	}
%\ea
%
%
%%%%%%%%%%%%%%%%%%%%%
%\be
%R_{\m\n}R^{\m\n} = 12\le(\f{\dot{a}^4}{a^8} - \f{\dot{a}^2\ddot{a}}{a^7}	+ \f{\ddot{a}^2}{a^6}\ri). \la{eq:Ricci-sq	}
%\ee
%
%%%%%%%%%%%%%%%%%%%%%
%%\be
%%R_{\m\n}R^{\m\n} = 12\le(\f{\dot{a}^4}{a^8} - \f{\dot{a}^2\ddot{a}}{a^7}	+ \f{\ddot{a}^2}{a^6}\ri). \la{eq:Ricci-sq	}
%%\ee
%
%
%\be
%\mbox{Gauss-Bonnet scalar:}~~G = 24\, \f{a\dot{a}^2\ddot{a} - \dot{a}^4}{a^8}.
%\ee
%
%%%%%%%%%%%%%%%%%%
%\be
%\square R = 6 \le(\f{3\ddot{a}^2}{a^6} - \f{6\dot{a}^2\ddot{a}}{a^7} + \f{4\dot{a}\dddot{a}}{a^6} - \f{\ddddot{a}}{a^5}\ri), \la{eq:boxR}
%\ee

\section{$s_q$, $u_q$ and $\t_q$ of different adiabatic orders} \la{app:sut}

\ba
\t_q^{(3)} = \f{1}{16\O_q^3} \le[-15 \f{\dot{\O}_q^3}{\O_q^3} + 4 \dot{Q} + 14\f{\dot{\O}_q \ddot{\O}_q}{\O_q^2} - 12 \f{\dot{\O}_q}{\O_q} Q - 2 \f{\dddot{\O}_q}{\O_q}\ri] \la{eq:t3}
\ea
%%%%%%%%%%%%%%%%%%%%%%%%%%%%
\ba
u_q^{(4)} &=& \f{1}{32\O_q^4}\le[9 \f{\dot{\O}_q^4}{\O_q^4} - 115 \f{\dot{\O}_q^2 \ddot{\O}_q}{\O_q^4} + 4(2Q^2 + \ddot{Q}) + 2\le(27 \f{\dot{\O}_q^2}{\O_q^2}Q + 7\f{\ddot{\O}_q^2}{\O_q^2} + 11 \f{\dot{\O}_q \dddot{\O}_q}{\O_q^2}\ri) \ri. \nn \\ 
 && \hs{1.5cm} \le. - 2\le(12\f{\dot{\O}_q}{\O_q} \dot{Q} + 8\f{\ddot{\O}_q}{\O_q}Q + \f{\ddddot{\O}_q}{\O_q} \ri) \ri]  \la{eq:u4}
\ea
%%%%%%%%%%%%%%%%%%%%%%%%%%%%
\ba
s_q^{(4)} &=& \f{1}{256\O_q^4}\le[- 45\f{\dot{\O}_q^4}{\O_q^4} + 16Q^2 + 40 \f{\dot{\O}_q^2 \ddot{\O}_q}{\O_q^4} +  16 \le(\f{\dot{\O}_q}{\O_q} \dot{Q} - \f{\ddot{\O}_q}{\O_q} Q\ri) \ri. \nn \\ 
 && \hs{1.5cm} \le. - 4 \le(\f{\dot{\O}_q^2}{\O_q^2} Q - \f{\ddot{\O}_q^2}{\O_q^2} + 2 \f{\dot{\O}_q\dddot{\O}_q}{\O_q^2} \ri) \ri]  \la{eq:s4}
\ea
%%%%%%%%%%%%%%%%%%%%%%%%%%%%
\ba
\t_q^{(5)} &=& \f{1}{256\O_q^5} \le[2835 \f{\dot{\O}_q^5}{\O_q^5} - 4620 \f{\ddot{\O}_q\dddot{\O}_q}{\O_q^4} + 140 \f{\dot{\O}_q}{\O_q} \le(10 \f{\dot{\O}_q^2}{\O_q^2}Q + 9\f{\ddot{\O}_q^2}{\O_q^2} + 7\f{\dot{\O}_q\dddot{\O}_q}{\O_q^2}\ri)  \ri. \nn \\ 
 && \hs{1.5cm} \le. - 16 \le(6Q\dot{Q} + \dddot{Q}\ri) - 8 \le(85\f{\dot{\O}_q^2}{\O_q^2}\dot{q} + 25\f{\ddot{\O}_q\dddot{\O}_q}{\O_q} + 110 \f{\dot{\O}_q\ddot{\O}_q}{\O_q}Q + 16\f{\dot{\O}_q\ddddot{\O}_q}{\O_q}\ri)  \ri. \nn \\ 
 && \hs{1.5cm} \le.  + 8 \le(30\f{\dot{\O}_q}{\O_q}Q^2 + 20 \f{\ddot{\O}_q}{\O_q}\dot{Q} + 20\f{\dot{\O}_q}{\O_q}\ddot{Q} + 10 \f{\dddot{\O}_q}{\O_q}Q + \f{\O_q^{(5)}}{\O_q} \ri) \ri] \la{eq:t5}
\ea
%%%%%%%%%%%%%%%%%%%%%%%%%%%%
\ba
u_q^{(6)} &=& \f{1}{512\O_q^6}\le[- 28350 \f{\dot{\O}_q^6}{\O_q^6} + 55755 \f{\dot{\O}_q^4 \ddot{\O}_q}{\O_q^6} + 70 \f{\dot{\O}_q^2}{\O_q^2} \le(169 \f{\dot{\O}_q^2}{\O_q^2}Q + 342 \f{\ddot{\O}_q^2}{\O_q^2} + 178\f{\dot{\O}_q\dddot{\O}_q}{\O_q^2}\ri)  \ri. \nn \\ 
 && \hs{1.5cm}  + 28 \le(220 \f{\dot{\O}_q^3}{\O_q^3}\dot{Q} + 45 \f{\ddot{\O}_q^3}{\O_q^3} + 210 \f{\dot{\O}_q\ddot{\O}_q\dddot{\O}_q}{\O_q^3} + 400\f{\dot{\O}_q^2\ddot{\O}_q}{\O_q^3}Q + 67 \f{\dot{\O}_q^2\ddddot{\O}_q}{\O_q^3}\ri)  \nn \\ 
 && \hs{1.5cm}  - 16 \le(6Q^3 + 6 \dot{Q}^2 + 8Q\ddot{Q} + \ddddot{Q} \ri) + 8 \le(230\f{\dot{\O}_q^2}{\O_q^2}Q^2 + 205 \f{\dot{\O}_q^2}{\O_q^2}\ddot{Q} + 25\f{\dddot{\O}_q^2}{\O_q^2} \ri.\nn \\ 
 && \hs{1.5cm} \le.  + 125 \f{\ddot{\O}_q^2}{\O_q^2}Q + 190\f{\dot{\O}_q\dddot{\O}_q}{\O_q^2}Q + 41\f{\ddot{\O}_q\ddddot{\O}_q}{\O_q^2}  + 400\f{\dot{\O}_q\ddot{\O}_q}{\O_q^2}\dot{Q} + 22\f{\dot{\O}_q \O_q^{(5)}}{\O_q^2}\ri)  \nn \\ 
 &&  \le. + 8\le(50 \f{\ddot{\O}_q}{\O_q}Q^2 + 40\f{\ddot{\O}_q}{\O_q}\ddot{Q} + 30\f{\dddot{\O}_q}{\O_q}\dot{Q} + 30\f{\dot{\O}_q}{\O_q}\dddot{Q} + 140\f{\dot{\O}_q}{\O_q}Q\dot{Q} + 12 \f{\ddddot{\O}_q}{\O_q}Q + \f{\O_q^{(6)}}{\O_q}\ri) \ri]  \la{eq:u6}
\ea
%%%%%%%%%%%%%%%%%%%%%%%%%%%%
\ba
s_q^{(6)} &=& \f{1}{2048\O_q^6}\le[4725\f{\dot{\O}_q^6}{\O_q^6} -7560 \f{\dot{\O}_q^4 \ddot{\O}_q}{\O_q^6} + 28 \f{\dot{\O}_q^2}{\O_q^2} \le(44\f{\dot{\O}_q^2}{\O_q^2}Q + 63 \f{\ddot{\O}_q^2}{\O_q^2} + 62\f{\dot{\O}_q\dddot{\O}_q}{\O_q^2}\ri)\ri. \nn \\ 
 && \hs{1.5cm} \le.  - 32 \le(4Q^3 - \dot{Q}^2 + 2Q\ddot{Q}\ri) + 32 \le(10 \f{\ddot{\O}_q}{\O_q}Q^2 + \f{\ddot{\O}_q}{\O_q}\ddot{Q} - \f{\dddot{\O}_q}{\O_q}\dot{Q} - \f{\dot{\O}_q}{\O_q}\dddot{Q} + \f{\ddddot{\O}_q}{\O_q}Q \ri) \ri. \nn \\  
 && \hs{1.5cm}  - 112 \le(11\f{\dot{\O}_q^3}{\O_q^3}\dot{Q} - \f{\ddot{\O}_q^3}{\O_q^3} + 3 \f{\dot{\O}_q\ddot{\O}_q\dddot{\O}_q}{\O_q^3} - \f{\dot{\O}_q^2\ddot{\O}_q}{\O_q^3}Q + 2\f{\dot{\O}_q^2\ddddot{\O}_q}{\O_q^3}\ri)  + 8 \le(30\f{\dot{\O}_q^2}{\O_q^2}Q^2 \ri.\nn \\ 
 && \hs{0.5cm} \le. \le.  - 32\f{\dot{\O}_q^2}{\O_q^2}\ddot{Q} - \f{\dddot{\O}_q^2}{\O_q^2}  + 44 \f{\ddot{\O}_q^2}{\O_q^2}Q + 12\f{\dot{\O}_q\dddot{\O}_q}{\O_q^2}Q + 2\f{\ddot{\O}_q\ddddot{\O}_q}{\O_q^2} - 44\f{\dot{\O}_q\ddot{\O}_q}{\O_q^2}\dot{Q} - 2\f{\dot{\O}_q \O_q^{(5)}}{\O_q^2}\ri)\ri]  \la{eq:s6}
\ea

%\begin{thebibliography}{99}
%
%

\end{document}